\documentclass[aps,prc,amsmath,amssymb,twocolumn,showpacs]{revtex4-1} 
\usepackage{epsfig} 
\usepackage{bm} 
\usepackage{graphicx}

\usepackage{amsmath}

\begin{document}

\title{Probing mixed-spin pairing in heavy nuclei}

\author{Brendan Bulthuis and Alexandros Gezerlis} 
\affiliation{Department of Physics, University of Guelph, Guelph, Ontario, N1G 2W1, Canada}

\date{\today}

\def\be{\begin{equation}}
\def\ee{\end{equation}}
\def\rb{\rangle}
\def\lb{\langle}
\def\half{\hbox{\rm${1\over 2}$}}

\newcommand{\ket}[1]{\left| #1 \right>} 
\newcommand{\bra}[1]{\left< #1 \right|} 

\begin{abstract}
The nature of the nuclear pairing condensate is an active topic of investigation, especially as regards its neutron-proton
versus identical-particle character, which manifests as the difference between spin-singlet
and spin-triplet pairing. In this work, we probe the recently proposed mixed-spin pairing condensates, using a 
phenomenological Hamiltonian and Hartree-Fock-Bogoliubov theory along with the gradient method. In addition to improving
the solution of the many-body problem, we have calculated a series of physical quantities and examined the robustness of
the mixed-spin pairing state as the input Hamiltonian is modified. 
Overall, we find that even though 
the mixed-spin correlation energy is suppressed in comparison to earlier work, the new pairing behavior persists.
We also discuss the possibility of directly probing the mixed-spin pairing phase.
\end{abstract}

\pacs{21.10.-k, 21.30.Fe, 21.60.Jz, 27.60.+j} 
 
\maketitle

\section{Introduction}

The interaction between two neutrons is nearly strong enough to produce a bound state, a situation that leads to
fruitful interplay between the physics of neutron-rich matter and that of ultracold atomic gases.\cite{Gandolfi}
``Nearly'' means that while the neutron-neutron scattering length is typically much larger than the average interneutron
spacing, it is not infinite: there is no bound dineutron. Similarly, there is no bound diproton. This, added to the existence
of the deuteron (a bound state of a neutron and a proton) led to the proposal of isospin-dependent interactions
and the conclusion
that the isospin-singlet, spin-triplet ($T=0, S=1$) interaction is stronger than the isospin-triplet, 
spin-singlet ($T=1, S=0$) one. 

While the isoscalar (spin-triplet) interaction is stronger than the isovector (spin-singlet) one, the pairing 
appearing in observed nuclei is found between identical particles (neutron-neutron and proton-proton). The question
of what changes in the transition from (two-particle) interaction to (many-particle) correlations has therefore been
actively investigated for a while now. A possible explanation is that the strong nuclear spin-orbit term suppresses
spin-triplet neutron-proton pairing more than it does spin-singlet identical-particle pairing. The concept of 
proton-neutron correlations has therefore been studied in a variety of settings, ranging from experiment, binding energy
systematics, shell-model, and mean-field pairing calculations for nuclei, 
\cite{Goodman1, Engel1, Engel2, 
SatulaWyss, Terasaki, Goodman3, PovesMartinez, Goodman4, Macchiavelli, 
Macchiavelli2,
Goodman2, Baroni,
Sagawa, Bai, vanIsacker,
Yoshida, Grodner, Hen, Bai2, Kanada, Fu, Sambataro, Gambacutra}
the study of (astrophysically 
relevant) nuclear matter, \cite{Garrido, Shulga} 
up to the possibility that proton-neutron mixing in the particle-hole sector is important.
\cite{Sheikh} 
A comprehensive and readable review on neutron-proton pairing has appeared recently. \cite{FrauendorfMacchiavelli}

Another reason why neutron-proton pairing might be disfavored (in comparison to identical-particle pairing)
is that most nuclei don't have the same number of protons and neutrons. This can be easily remedied by eliminating 
the isospin polarization, i.e. studying $N=Z$ nuclei (as is done in several of the references cited above). 
Combining the two possibilities (reduced effects of the spin-orbit field and no isospin polarization),
Ref.~\cite{BertschLuo} used a phenomenological interaction and Hartree-Fock-Bogoliubov (HFB) theory to study heavy nuclei
(where the surface-to-volume ratio is small) with $N=Z$, finding tantalizing hints of the possibility of spin-triplet
pairing, which were, however, above the proton dripline. Ref.~\cite{GezerlisBertschLuo} built on that work 
by examining $N \geq Z$ nuclei, finding spin-triplet-paired nuclei with unequal neutron and proton numbers, as well
as a smooth transition from spin-triplet to spin-singlet, which went through a region of ``mixed-spin pairing''. 
The latter nuclei extended below the proton dripline and were thus, in principle, accessible to experiment. (Intriguingly, a mixed pairing state has also been investigated in the context
of a one-dimensional Hubbard model~\cite{Sun}).

In the solution of the HFB equations, Refs. \cite{BertschLuo} and \cite{GezerlisBertschLuo} dropped the Hartree-Fock
$\Gamma$ term, assuming this was effectively absorbed into the model Hamiltonian. In the present work, after a 
brief summary of HFB theory (section~\ref{sec:HFB}), the gradient method (section~\ref{sec:grad}),
and our input Hamiltonian (section~\ref{sec:Ham}),
we return to this problem and explicitly include the $\Gamma$ field in our many-nucleon calculations 
(section~\ref{sec:Gamma}). We then explore the new term's effects on the pairing character for different nuclei 
(section~\ref{sec:pair}), while also probing particle-number fluctuations (section~\ref{sec:numres})
and the transition from spin-triplet to spin-singlet in more detail (section~\ref{sec:cont}). We proceed
to explore possible experimental signatures of the spin-triplet and mixed-spin pairing states (section~\ref{sec:gaps}).
We conclude by examining the dependence on the one-body (section~\ref{sec:one}) and two-body (section~\ref{sec:two}) 
parameters in our input model Hamiltonian.

\section{Formalism}

\subsection{Hartree-Fock-Bogoliubov theory}
\label{sec:HFB}

In this section we briefly go over the basics of Hartree-Fock-Bogoliubov theory \cite{RingSchuck}, mainly
in order to establish the notation that will be used throughout the paper.
HFB theory starts from the Fock-space representation of the Hamiltonian:
\begin{equation}
\label{H}
\hat H = \sum_{ij} \varepsilon_{ij} c^\dagger_i c_j + \frac{1}{4} \sum_{ijkl} \bar{v}_{ijkl}c^\dagger_i
c^\dagger_j c_l c_k
\end{equation}
Then, employing a general Bogoliubov transformation one introduces quasiparticle operators $\beta_i^{\dagger}$ and $\beta_i$:
\begin{equation}
\beta_i^{\dagger} = \sum_{j} \left ( U_{ji} c_j^{\dagger} + V_{ji}c_j \right )
\end{equation}
This already introduces the basic $U$ and $V$ variables, which often appear in matrix form as below:
\begin{equation}
\left( \begin{array}{c}
\beta \\
\beta^\dagger 
\end{array}
\right) =
\left[ \begin{array}{cc}
U^\dagger & V^\dagger \\
V^T & U^T
\end{array}
\right]
\left( \begin{array}{c}
c \\
c^\dagger
\end{array}
\right)\,
\end{equation}
The Hamiltonian can then be re-expressed in the quasiparticle basis as follows:
\begin{equation}
\hat H = H^{00} + \beta^{\dagger}  H^{11} \beta  + \frac{1}{2} \beta^{\dagger} H^{20}
\beta^{\dagger} + \cdots
\end{equation}
where the superscripts have the obvious meaning of counting the numbers of the quasiparticle creation and
annihilation operators (the $...$ represents $H^{40}$, $H^{31}$, $H^{22}$ and hermitian conjugates).

HFB theory amounts to using a variational principle and HFB wave functions $| \Phi \rangle$ 
(along with Thouless' theorem) 
to minimize the expectation value of $\hat{H}$, and then writing:
\begin{equation}
\hat H = H^{00} + \sum_k E_k \beta^\dagger_k  \beta_k
\label{eq:quasiH}
\end{equation}
where the $E_k$ are known as quasiparticle energies and are the eigenvalues in the following
HFB equations:
\begin{equation}
\left[ \begin{array}{cc}
h'  & \Delta \\
-\Delta^* & -{h'}^* 
\end{array}
\right]
\left( \begin{array}{c}
U_k\\
V_k
\end{array}
\right)
=
\left( \begin{array}{c}
U_k\\
V_k
\end{array}
\right) E_k\,.
\end{equation}
where $U_k$, $V_k$ are the columns of the matrices $U$, $V$. The minimization takes place subject
to certain constraints, such as fixed proton and neutron numbers. Thus, 
$h' = h - \sum \lambda_i Q_i$, where $Q_i$ are constraining fields to be discussed below and $\lambda_i$ 
are Lagrange multipliers. The $h$ itself consists of the single-particle contribution along with a self-consistent
field $\Gamma$ known from Hartree-Fock theory: $h = \varepsilon + \Gamma$. 
We thus see that the solution of these equations
rests upon the knowledge of the $\Gamma$ and $\Delta$ entities:
\begin{align}
\Gamma_{ij} &= \sum_{kl}\bar{v}_{iljk}\rho_{kl} \nonumber \\
\Delta_{ij} &= \frac{1}{2}\sum_{kl}\bar{v}_{ijkl}\kappa_{kl}
\label{eq:GD}
\end{align}
where the pairing field $\Delta$ encapsulates the physics relating to a superfluid state.
The $\Gamma$ and $\Delta$, in their turn, are given in terms of the normal 
density $\rho$ and the anomalous density $\kappa$:
\begin{align}
\rho &= V^*V^T \nonumber \\
\kappa &= V^*U^T
\end{align}
Note that $\rho$ is Hermitian, $\kappa$ is skew symmetric, and that $\rho$ and $\kappa$ together 
uniquely determine the wave function $| \Phi \rangle$.

We see from the Hamiltonian in the quasiparticle basis, Eq.~(\ref{eq:quasiH}), that the quasiparticle vacuum, 
$| \Phi \rangle$, has the eigenvalue $H^{00}$ which turns out to have the form:
\begin{equation}
H^{00} = Tr\left(\varepsilon\rho +\frac{1}{2}\Gamma\rho -\frac{1}{2}\Delta\kappa^*\right).
\label{eqn_HamExpect}
\end{equation}
This contains contributions from: a) the single-particle $\varepsilon$ term, b) the Hartree-Fock $\Gamma$ 
field, and c) the pairing field $\Delta$. 
The task of HFB theory can be viewed as the minimization of this $H^{00}$ subject to constraints. 
It is straightforward to see how intimately connected to the $U$ and $V$ matrices the solution of the entire problem is.

\subsection{Gradient Method}
\label{sec:grad}
Minimizing the $H^{00}$ subject to constraints can be a daunting task, especially when several constraints are involved.
The simplest case of including constraints relates to calculations which try to constrain the average particle number. In the present work, as in Ref.~\cite{GezerlisBertschLuo}, we employ a number of constrained parameters to fully investigate different types of pairing condensates, and hence require an efficient method to solve the HFB problem. Here we give a
flavor of the principles involved and refer to Refs.~\cite{RingSchuck,Egido,RobledoBertsch} for further details.

The gradient method begins with a trial wavefunction $\left|\Phi_0\right>$, defined by the matrices $U_0$ and $V_0$, and uses Thouless' theorem to define a wavefunction dependent on an antisymmetric matrix $Z$, $\left|\Phi(Z)\right>$:
\begin{equation}
\ket{\Phi(Z)} = N ~\text{exp}\left(\sum_{i<j}Z_{ij}\beta^\dagger_i\beta^\dagger_j\right)\ket{\Phi_0},
\end{equation}
where $N$ is a normalization factor, $Z$ is the antisymmetric Thouless matrix, and $\beta_i$ annihilates $\ket{\Phi_0}$.
The energy expectation value, to first order in $Z$, is then found to be:
\begin{eqnarray}
E(Z)&=&\bra{\Phi(Z)}\hat{H}\ket{\Phi(Z)}\nonumber\\
&=&H^{00}+\sum_{i<j}\left(Z^{*}_{ij}H^{20}_{ij}+Z_{ij}H^{20*}_{ij}\right).
\label{eqn_energyWithZ}
\end{eqnarray}
Taking a derivative with respect to $Z^*$ of the energy expectation value with this new wavefunction gives a new antisymmetric matrix $Z_1$, namely
\begin{equation}
Z_{1ij} = -\eta\frac{\partial}{\partial Z^*_{ij}}\left<\Phi(Z)\right|\hat{H}\left|\Phi(Z)\right> = -\eta H^{20}_{ij}.\label{eqn_newZ}
\end{equation}
The parameter $\eta$ is somewhat arbitrary, and determines how large the step is.
This $Z_1$ matrix is then used to transform the $U_0$ and $V_0$ into new matrices $U_1$ and $V_1$:
\begin{eqnarray}
U_{1}&=&U_{0}+V^{*}_{0}Z^{*}_{1}\nonumber\\
V_{1}&=&V_{0}+U^{*}_{0}Z^{*}_{1}.
\end{eqnarray}
which in turn define a new wavefunction $\left|\Phi_1\right>$. This process is repeated until 
self-consistency, when $H^{20}$ is zero, as in Eq.~(\ref{eq:quasiH}).

Adding in constraints is relatively simple.
We define $\hat{Q}_i$ to be the operator corresponding to the $i$th constraint, which we give a value of $Q_i$. By also defining Lagrange multipliers $\lambda_i$, we can include these constraints by adding them into the Hamiltonian:
\begin{equation}
\hat{H} \rightarrow \hat{H} - \sum_i\lambda_i\hat{Q}_i.
\end{equation}
This results in the $Z_1$ matrix becoming:
\begin{equation}
Z_1 = -\eta\left(H^{20}-\sum_i\lambda_i{Q}^{20}_i\right).
\label{eq:Z1}
\end{equation}
In order to determine what the $\lambda_i$ are, we impose the condition that the expectation of $\hat{Q}_i$ equals $Q_i$, to first order in $Z$. The operators also have their own quasiparticle basis representations, similar to that of the Hamiltonian, with their corresponding $Q^{00}_i$, $Q^{11}_i$, and $Q^{20}_i$, so they have an expectation value of
\begin{equation}
\left<\hat{Q}_i\right> = Q_i^{00}+\sum_{k<l}\left(Z^*_{1,kl}Q^{20}_{i,kl}+Z_{1,kl}Q^{20*}_{i,kl}\right)=Q^{00}_i+Z_1\cdot Q^{20}_i.
\end{equation}
Our condition then requires $Z_1\cdot Q^{20}_i$ to be zero. Using our new $Z_1$ 
from Eq.~(\ref{eq:Z1}) with the included constraints then leads to a system of equations involving the $\lambda_i$:
\begin{equation}
\sum_j\lambda_jQ^{20}_j\cdot Q^{20}_i = H^{20}\cdot Q^{20}_i.
\end{equation}
This allows us to relatively easily solve for $\lambda_i$.
Once these $\lambda_i$'s are determined, this $Z_1$ is used as normal. In the present work, up to 8 constraining fields are explored, 2 for the neutron and proton particle numbers and 6 for the pairing configuration amplitudes.

\subsection{Particle Number Fluctuations}
\label{sec:num}

One can also use the above formalism to examine the question of particle-number fluctuations in this theory.
\cite{RobledoBertsch}
For a single-particle operator $\hat{O}$:
\begin{equation}
O_{ij}c^{\dagger}_{i}c_{j}=O^{00}+O^{11}_{ij}\beta^{\dagger}_{i}\beta_{j}+\frac{1}{2}\left(O^{20}_{ij}\beta^{\dagger}_{i}\beta^{\dagger}_{j}+O^{02}\beta_{i}\beta_{j}\right),
\end{equation}
we can calculate the statistical variance
\begin{equation}
\langle\Delta \hat{O}^2\rangle = \langle \hat{O}^2 \rangle - \langle \hat{O} \rangle^2
\end{equation}
reformulated in the HFB formalism, using the ground state $\ket{\Phi}$. Thus, the variance becomes
\begin{eqnarray}
\langle \Delta \hat{O} \rangle^2 &=& \frac{1}{2}\text{Tr}\left(O^{20}O^{02}\right).
\label{eqn_variance}
\end{eqnarray}
Below we will use separate operators for the proton and neutron numbers and will thereby determine the 
proton and neutron number fluctuations (defined as the square root of the variance).

\subsection{Hamiltonian}
\label{sec:Ham}
We consider a phenomenological Hamiltonian, consisting, 
as usual, of a one-body part and a two-body interaction term, represented as
\begin{align}
\hat{H} &= \sum_{ij} \varepsilon_{ij} c_i^\dagger c_j +\frac{1}{2} \sum_{ijkl} v_{ijkl} c^\dagger_i c^\dagger_j c_l c_k 
\nonumber \\
v_{ijkl} &= \int d^3rd^3r'\chi^{*}_{i}(\mathbf{r})\chi^{*}_{j}(\mathbf{r}')V(\mathbf{r},\mathbf{r}')\chi_{k}(\mathbf{r})\chi_{l}(\mathbf{r}') 
\label{eqn_Ham}
\end{align}
the only difference from Eq.~(\ref{H}) being that now we are not using antisymmetrized matrix elements. The $\chi_i(\mathbf{r})$ are coordinate-space wave functions associated with a single-particle Hamiltonian.

In order to probe the essential features of pairing in heavy nuclei, we here follow Refs.~\cite{BertschLuo,GezerlisBertschLuo}
and take the one-body part $\varepsilon$ to contain a kinetic energy, a Woods-Saxon well, and a spin-orbit term:
\begin{align}
\varepsilon &= \frac{p^2}{2m}+V_{WS}f(r) + \mathbf{l}\cdot{\mbox{\boldmath$\sigma$}}\hspace{1mm}V_{SO}\frac{1}{r}\frac{df(r)}{dr}\nonumber \\
f(r) &= \frac{1}{1+e^{(r-R)/a}} 
\label{eqn_WSfunc}
\end{align}
where $R$ is the radius of the nucleus, $R \approx 1.27 A^{1/3}$. The other parameters are taken to be:
$a = 0.67$ fm, $V_{WS} = -50$ MeV, and $V_{SO} = 33$ MeV fm$^2$, \cite{BertschLuo,GezerlisBertschLuo} 
though these are also probed in more detail in 
a later section.

Similarly, the two-body interaction term is taken to be of contact form:
\begin{align}
V(\mathbf{r},\mathbf{r}') &= \sum_{\alpha=1}^6 v_\alpha \delta^{(3)}(\mathbf{r}-\mathbf{r}')P_{L=0}P_{\alpha} \nonumber \\
&= \frac{1}{4} \bigg ( 3v_t+v_s
+(v_t-v_s){\mbox{\boldmath$\sigma$}}\cdot{\mbox{\boldmath$\sigma$}}') \nonumber \\
&\quad \times
\delta^{(3)}(\mathbf{r}-\mathbf{r}')P_{L=0}
 \bigg )
\label{eqn_interaction}
\end{align}
The operator $P_{L=0}$ projects onto states with zero total orbital angular momentum. This restriction means nuclear deformations are not considered, but it simplifies the calculations involved by imposing a block-diagonal 
structure on the $U$ and $V$ matrices. 
The index $\alpha$ represents one of the six pairing configurations shown in Table~\ref{Table}, which are explicitly
taken into account when we construct the pairing condensates. The 
operator $P_\alpha$ projects onto a particular pairing configuration. Obviously, the pairing strengths $v_\alpha$ 
can take on 
only two values, $v_s$ and $v_t$ for spin-singlet and spin-triplet, respectively, as explicitly shown in the second step of Eq.~(\ref{eqn_interaction}). 
These are taken to have values of 300 MeV and 450 MeV, respectively, to begin with (having been fit to shell-model
matrix elements) though their values will be varied in 
a later section.
Finally, the Dirac delta interaction is smeared by including only orbitals with single-particle energies $\pm 5$ MeV 
away from
the Fermi energy.

\begin{table}[t]
\begin{center}
\begin{tabular}{|c|cccccc|}
\hline
$\alpha$ & 1& 2& 3& 4& 5& 6\\
\hline
$(S,S_z)$ &(0,0)&(0,0) &(0,0) &(1,1) &(1,0)&(1,-1)\\
$(T,T_z)$  &(1,1) &(1,0)&(1,-1)&(0,0)&(0,0) &(0,0)\\
\hline
\end{tabular}
\label{Table}
\end{center}
\caption{Pairing channels organized by spin-isospin. $S$ is the two-nucleon spin, $S_z$ its $z$-projection;
$T$ is the two-nucleon isospin, $T_z$ its $z$-projection.
}
\end{table}

\subsection{Implementation of $\Gamma$ and $\Delta$}
\label{sec:Gamma}

In Refs.~\cite{BertschLuo,GezerlisBertschLuo}, given that the Hamiltonian was phenomenological, the $\Gamma$ matrix was assumed to be implicitly included in order to simplify the calculations. In this work, we explicitly introduce the 
$\Gamma$ matrix corresponding to the specific interaction described in the previous section.

If we introduce the interaction of Eq.~(\ref{eqn_interaction}) into the expression for the two-body matrix elements
in Eq.~(\ref{eqn_Ham}), we can express things more compactly:
\begin{equation}
v_{ijkl}= \sum_{\alpha=1}^6 v_\alpha\int d^3rd^3r'\delta^{(3)}(\mathbf{r}-\mathbf{r}')\langle ij| \mathbf{r},\mathbf{r}'\rangle P_{\alpha}P_{L=0} \langle \mathbf{r},\mathbf{r'} | kl \rangle.
\end{equation}
The kets $\ket{ij}$ contain all spatial, angular momentum, spin, and isospin quantum numbers. 
We can thus separate them into a spatial ket (which $P_{\alpha}$ does not affect) 
and another ket containing the remaining quantum numbers (which $P_{\alpha}$ does affect). 
Expressing the projection operator $P_{\alpha}$ as $\ket{\alpha}\bra{\alpha}$
allows us to introduce the following factors $A_{\alpha,ij}$, 
naturally defined in terms of Clebsch-Gordan coefficients:
\begin{eqnarray}
A_{\alpha,ij}&=&\sqrt{2}\left(\left.\frac{1}{2},s_{zi};\frac{1}{2},s_{zj}\right|S_\alpha S_{z\alpha}\right)\left(\left.\frac{1}{2},t_{zi};\frac{1}{2},t_{zj}\right|T_\alpha T_{z\alpha}\right)\nonumber\\
&&\times\left(-1\right)^{l_i-l_{zi}}\delta_{l_i,l_j}\delta_{l_{zi},l_{zj}}.
\end{eqnarray}
If we now integrate over $\mathbf{r'}$, 
write the spatial bra-kets in terms of single-particle wave functions, 
extract their spherically symmetric part, and carry out the angular integral we find:
\begin{equation}
\bar{v}_{ijkl}=\frac{1}{4\pi}\sum_{\alpha=1}^{6}v_\alpha\int drr^2\phi^{*}_{i}(r)\phi^{*}_{j}(r)\phi_{k}(r)\phi_{l}(r) A_{\alpha,ji}A_{\alpha,kl}.
\end{equation}
where the $\phi_i(r)$ are radial functions and we also switched to antisymmetrized 
matrix elements. 

It is now straightforward to calculate the $\Delta$ and $\Gamma$ fields, starting from Eq.~(\ref{eq:GD}).
For $\Delta$ we have:
\begin{align}
\Delta_{ij} &= \frac{1}{2}\sum_{kl}\bar{v}_{ijkl}\kappa_{kl} \nonumber \\
&= \sum_{\alpha=1}^6 v_\alpha\int dr r^2\phi^*_i(r)\phi^*_j(r)A_{\alpha,ij}\kappa_\alpha (r)
\end{align}
where
\begin{equation}
\kappa_\alpha (r) =\frac{1}{4\pi}\sum_{k<l}\phi_k(r)\phi_l(r)A_{\alpha,lk}\kappa_{kl}.
\end{equation}
as shown in the Appendix of Ref.~\cite{BertschLuo}. Similarly, we now find for the $\Gamma$ matrix:
\begin{align}
\Gamma_{ij} &= \sum_{kl}\bar{v}_{iljk}\rho_{kl} \nonumber \\
&= \sum_{\alpha kl}\frac{v_\alpha}{4\pi}\int dr r^2\phi^*_i(r)\phi^*_j(r)\phi_k(r)\phi_l(r)A_{\alpha,il}\rho_{lk}A_{\alpha,kj}.
\end{align}
For both the $\Delta$ and $\Gamma$ terms, we have written the final results in a form that highlights the 
matrix multiplications involved. Note that since there is no trace of a matrix multiplication (as 
in the case of the $\Delta$ matrix) there is no separation of steps needed for the $\Gamma$ matrix.

\section{Results}

\subsection{Pairing below the $N=Z$ line}
\label{sec:pair}

\begin{figure}[t]
\includegraphics[scale=0.45]{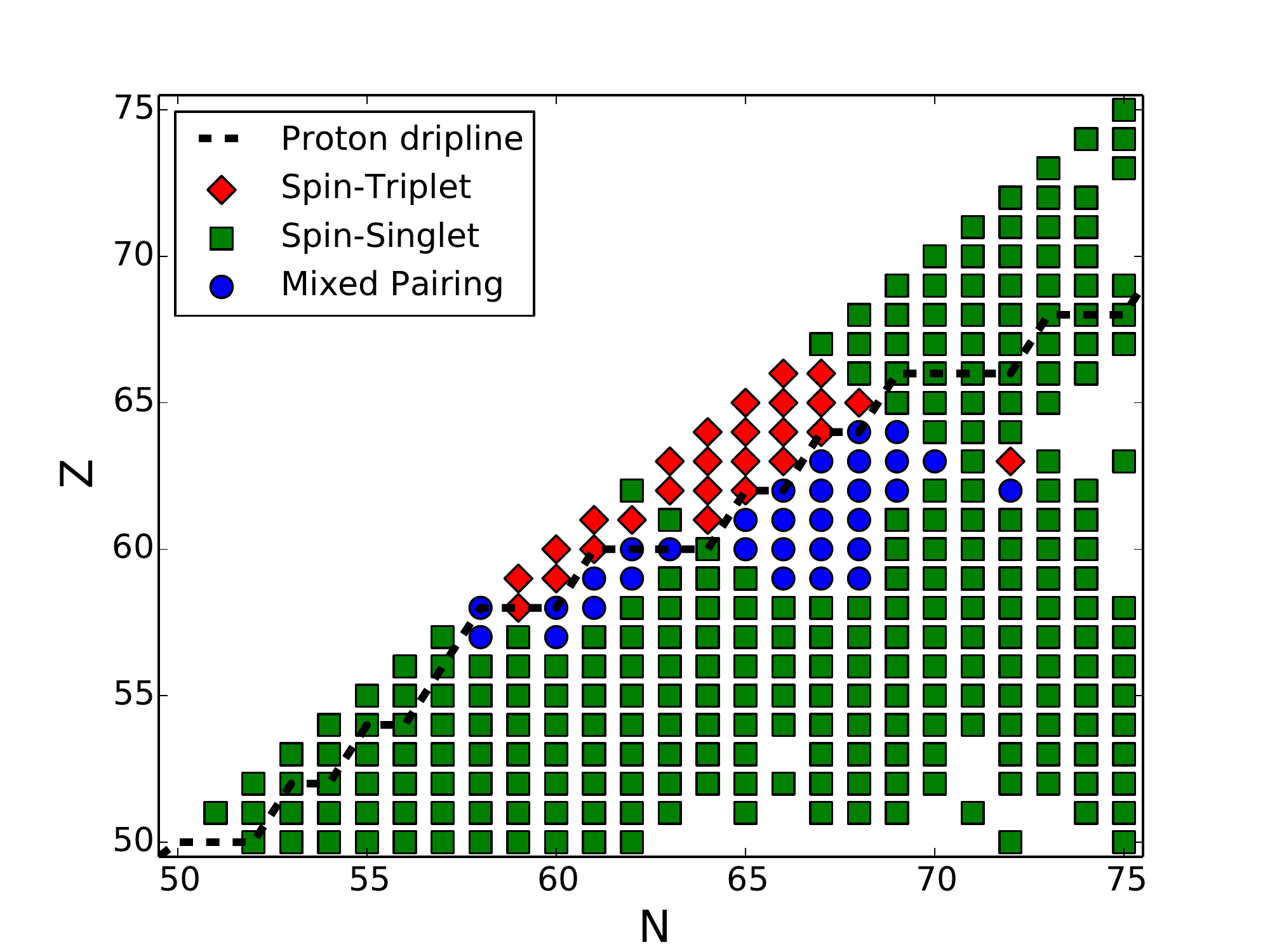}
\caption{(Color online) Chart of nuclides with $N \geq Z$ from $N=50$ to $N=75$. Blank squares indicate nuclei with no pairing ($E_{corr}<0.5$ MeV). Green squares denote a pairing condensate of mostly spin-singlet character, red diamonds those of a spin-triplet character, and blue circles indicate a mixture of the two types of pairing. 
\label{pair_fig}}
\end{figure}

In what follows, we will be interested in probing the effects of pairing in heavy nuclei. To do that, we
calculate the ground-state energy of a many-nucleon system when no pairing is present (calling that $E_0$), 
as well as the energy when pairing is turned on (calling that $E$). To ensure that $E_0$ corresponds to the case
where no pairing is present, we explicitly constrain all 6 pairing amplitudes ($\kappa_{\alpha} = \text{Tr}P_{\alpha}\kappa$)
to zero.
The difference between the two energies is the
\textit{correlation energy}:
\begin{equation}
E_{\text{corr}} = E_0 - E
\end{equation}
In Fig.~\ref{pair_fig}, nuclei below the $N=Z$ line have been mapped out for neutron numbers between 50 and 75. 
(Here and in the rest of this paper results shown include the $\Gamma$ term in the many-body calculations.)
Some 
nuclei have very small correlation energies, indicated by blank spaces, while most of the remaining nuclei exhibit spin-singlet pairing, denoted by green squares. Between neutron numbers 60 and 70 there is an island of nuclei that exhibit spin-triplet pairing (red diamonds), and another region of mixed pairing (blue circles). The mixed-spin pairing has been defined to be a state for which the spin-singlet amplitude is between one quarter and three quarters of the total pairing amplitude.
Also shown in the figure (black dashed line) is the proton dripline, produced based on Ref. \cite{Moller}.

\begin{figure*}[t]
\begin{minipage}[b]{0.46\linewidth}
\includegraphics[scale=0.65]{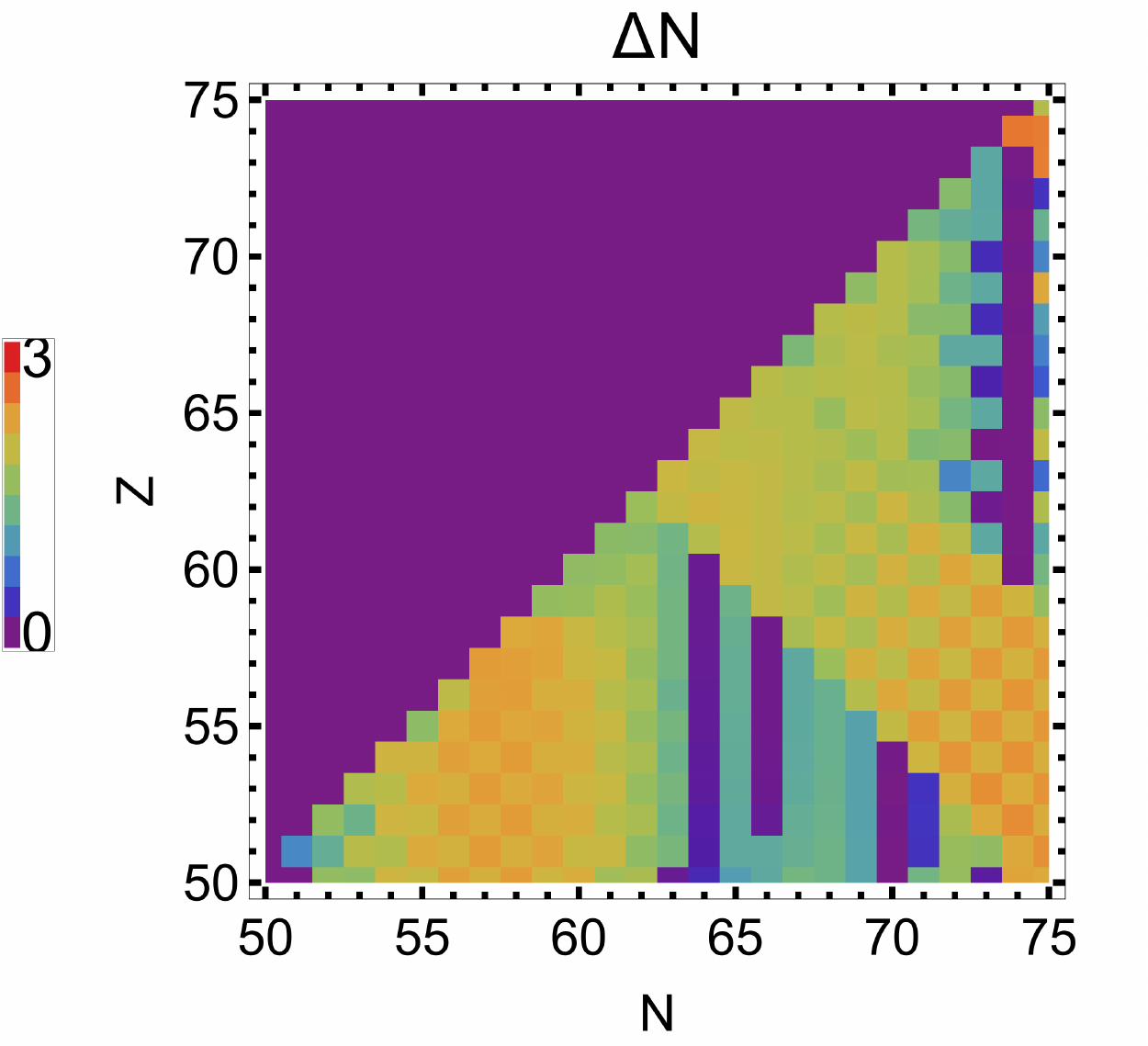}
\end{minipage}\hfill%
\begin{minipage}[b]{0.46\linewidth}
\includegraphics[scale=0.55]{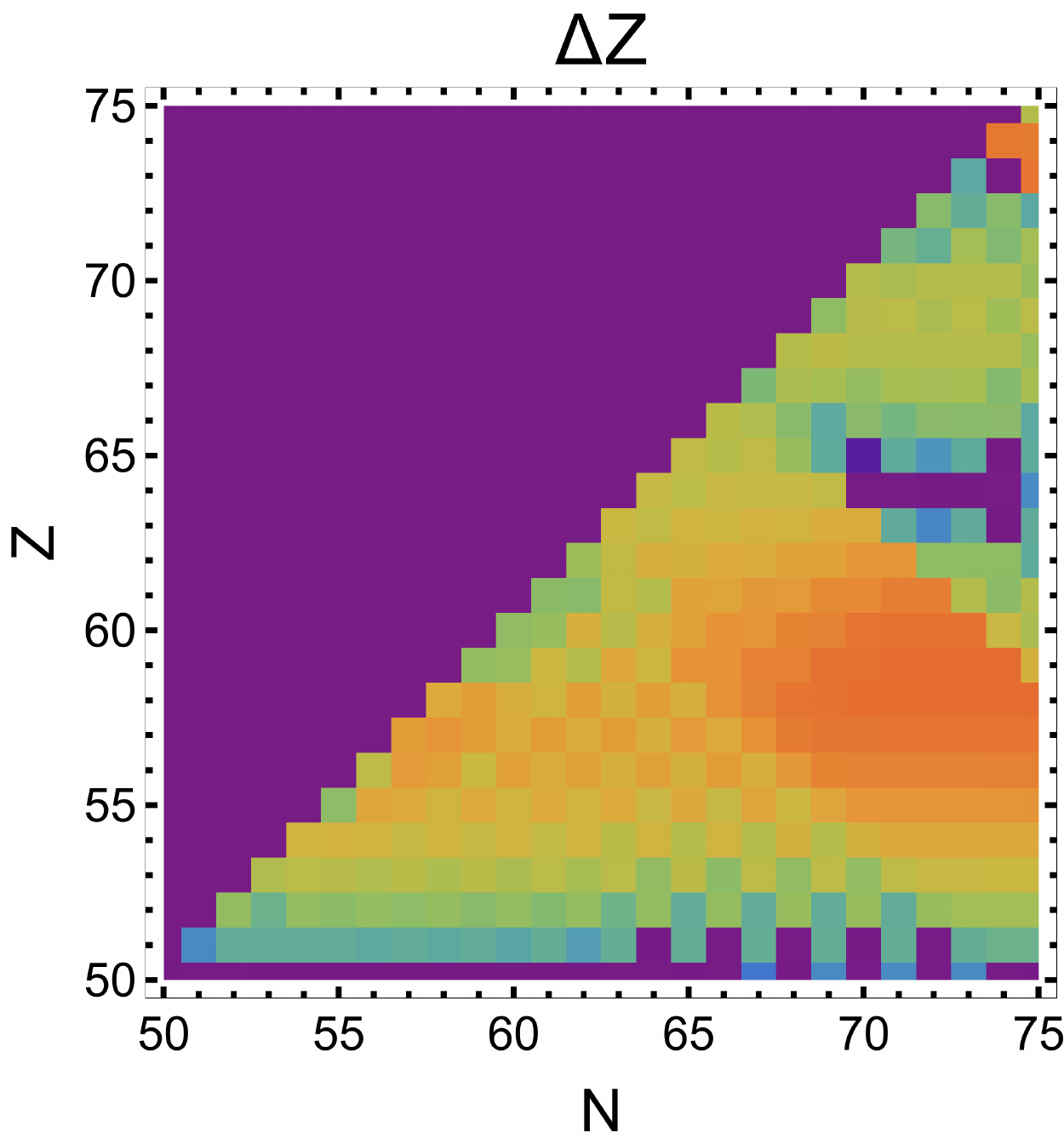}
\end{minipage}%
\caption{(Color online) Heat maps showing the fluctuations in neutron number (left panel) and proton number (right panel) for heavy nuclei on and off the $N=Z$ line. They range from 0 particles to a little over 3 particles. \label{fig_numFlux}}
\end{figure*}

Comparing this figure with Fig. 1 in Ref. \cite{GezerlisBertschLuo}, it can be seen that the inclusion of the $\Gamma$ matrix has a slight shifting or suppressing effect on spin-triplet pairing, as both the spin-triplet and mixed-spin regions have shifted closer to the $N=Z$ line. Nevertheless, the majority of the mixed-spin region remains below the proton-dripline, and thus may still be relevant to experiment. Detailed values of the correlation energy in the cases
of including and excluding the $\Gamma$ term will be discussed in connection with Fig.~\ref{fig_trans1} below.

\subsection{Number fluctuations}
\label{sec:numres}

Since HFB theory does not conserve particle number exactly, but only on average, it is worthwhile to investigate
the particle-number fluctuations. To do this, we have employed Eq.~(\ref{eqn_variance}), with separate operators 
for the proton and neutron numbers. The results are shown in Fig.~\ref{fig_numFlux}. Overall, we see that the 
fluctuations never exceed 4 particles (or 3\%). 

Two main features stand out in this figure. First, there are regions of the chart that have zero particle-number fluctuation.
As can be seen by comparing with the correlation energies that led to Fig.~\ref{pair_fig}, no fluctuations are trivially
equivalent to the absence of pairing, which automatically restores good particle number. There is a pretty consistent line
at the magic number of $Z=50$. Second, there are two areas of relatively higher fluctuations in both numbers, one centered at $(N,Z)=(70,60)$ and a smaller one at $(N,Z)=(57,55)$. As could have been expected, these closely
correlate with regions having large correlation energy.

\subsection{Singlet to triplet transition}
\label{sec:cont}
\begin{figure*}
\begin{minipage}[t]{0.3\linewidth}
\includegraphics[scale=0.45]{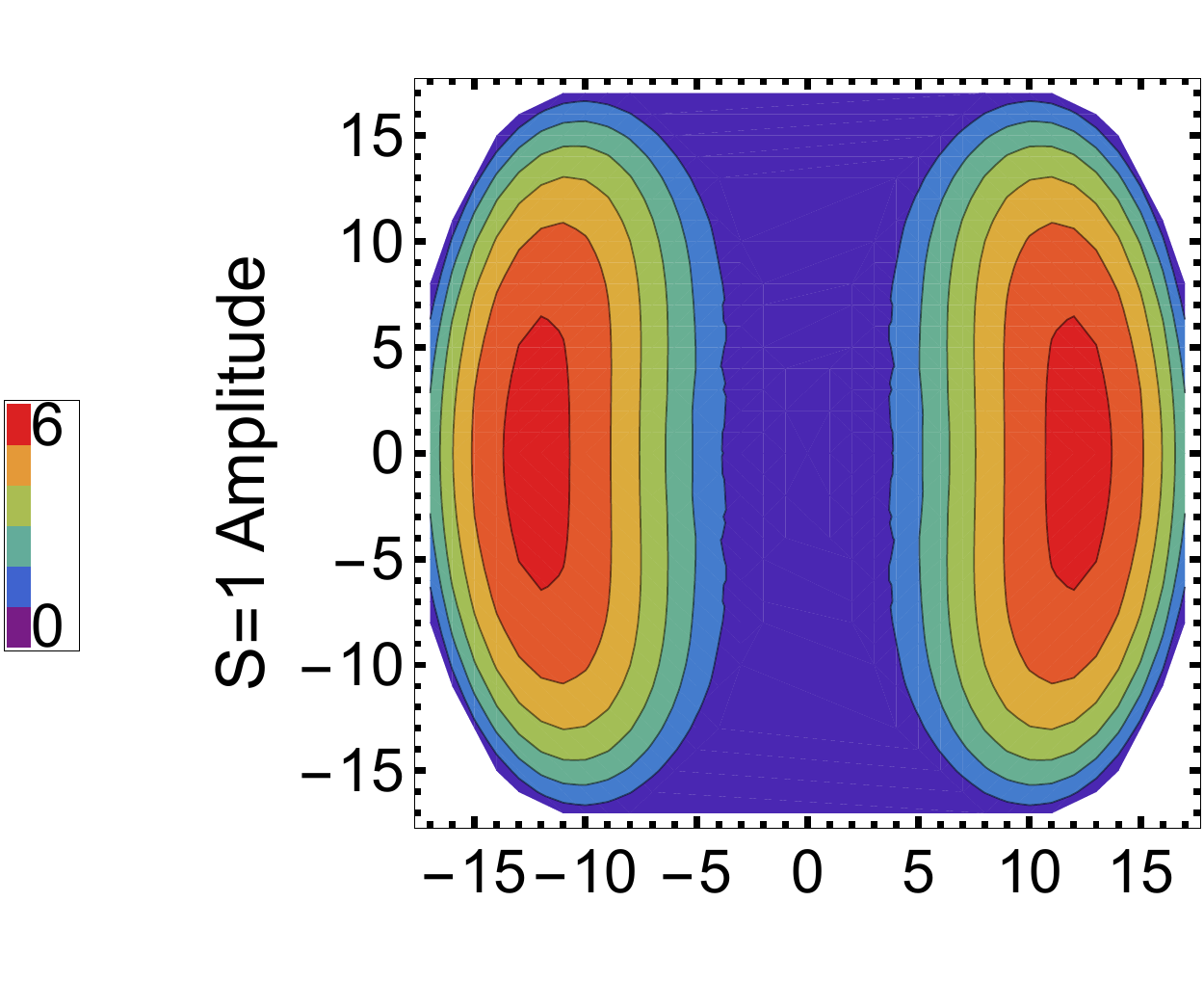}
\end{minipage}\hfill%
\begin{minipage}[t]{0.4\linewidth}
\includegraphics[scale=0.32]{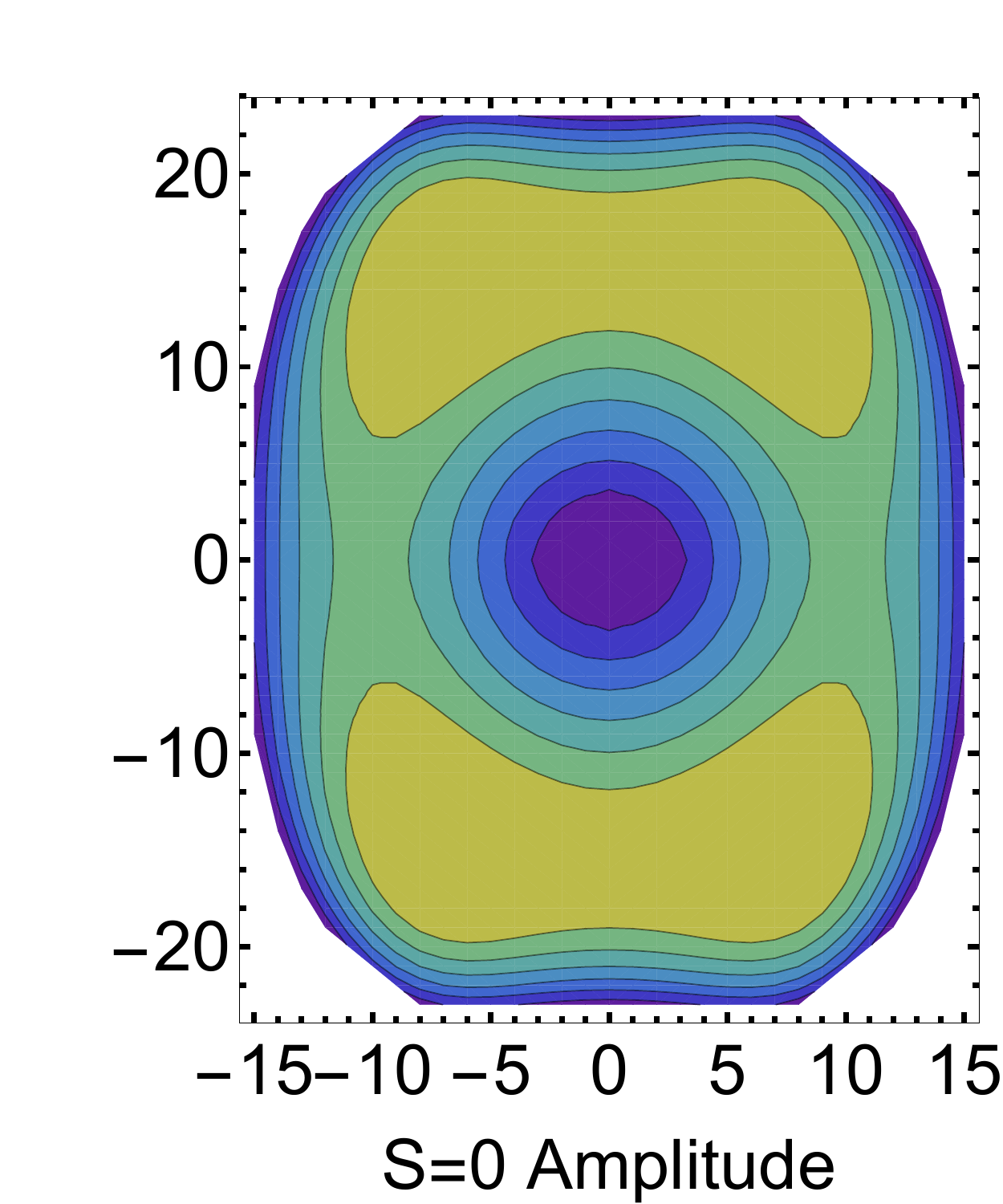}
\end{minipage}\hfill%
\begin{minipage}[t]{0.28\linewidth}
\includegraphics[scale=0.32]{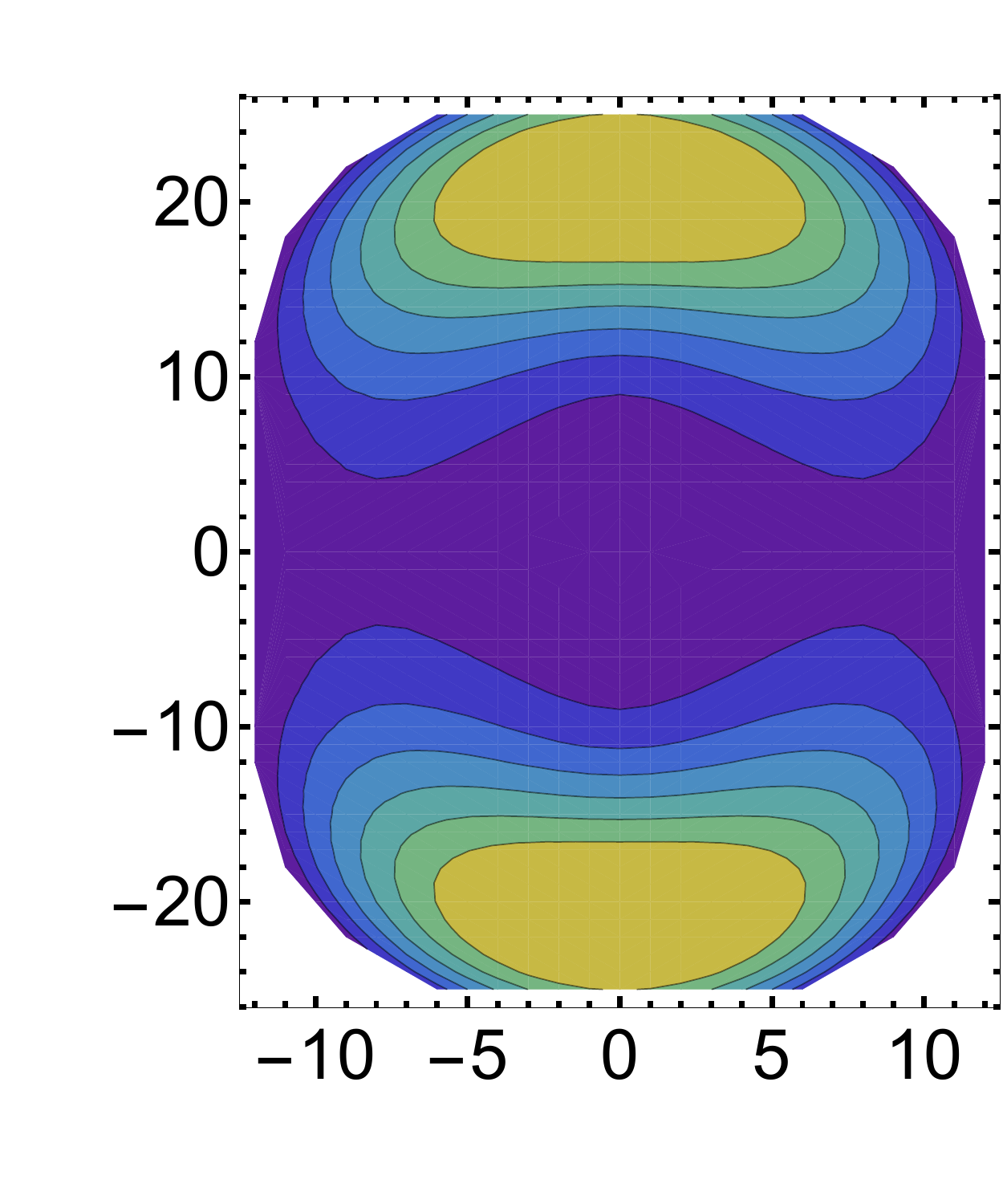}
\end{minipage}%
\caption{(Color online) Contour plots showing the correlation energy (in MeV) in 3 different nuclei on the $A=132$ line as a function of spin-singlet and spin-triplet pairing amplitudes. From left to right are $^{132}_{60}$Nd, $^{132}_{64}$Gd, and $^{132}_{66}$Dy. $^{132}_{60}$Nd exhibits mostly spin-singlet pairing, and $^{132}_{66}$Dy exhibits mainly spin-triplet pairing. $^{132}_{64}$Gd exhibits a mixed-state pairing that shows a smooth transition from spin-triplet to spin-singlet as $N-Z$ increases.  \label{fig_trans1}}
\end{figure*}

As in Ref.~\cite{GezerlisBertschLuo}, in order to more carefully examine the transition from spin-triplet to spin-singlet 
as a function of increasing $N-Z$, we study selected nuclei with $A=132$. This line was chosen as it exhibits 
examples of the three possible states: spin-singlet, spin-triplet, and mixed pairing condensates. Specifically, the three
nuclei were chosen to be  
$^{132}_{60}$Nd, $^{132}_{64}$Gd, and $^{132}_{66}$Dy. 
In addition to the ground-state results shown in Fig.~\ref{pair_fig},
we have here artificially constrained the spin-triplet and spin-singlet amplitudes away from the minima (explicitly: 
channels 5 and 1/3 from Table~\ref{Table}).
The resulting correlation energies are shown on a contour plot in Fig.~\ref{fig_trans1}.

The plot for $^{132}_{60}$Nd on the left has peaks at $(\pm 14, 0)$, indicating a spin-singlet condensate, which is the more
ordinary state of affairs. The unconstrained correlation
energy at the minimum is $\sim 6.5$ MeV, to be compared with the $\sim 8$ MeV 
when the $\Gamma$ term is excluded. 
The plot for $^{132}_{66}$Dy on the right shows peaks at $(0,\pm 22)$, indicating a spin-triplet condensate. Here the 
correlation energy is $\sim 4$ MeV, to be compared with $\sim 11$ MeV without the $\Gamma$ term. This is a major reduction, 
clearly reflecting the fact that the $\Gamma$ term serves to suppress the exotic spin-triplet state. Finally, 
the middle plot for $^{132}_{64}$Gd shows peaks at $(\pm 8,\pm 19)$, indicating a mixed-spin pairing condensate. 
In this case the correlation energy is $\sim 4$ MeV, to be compared with $\sim 7$ MeV without $\Gamma$, an intermediate reduction.
Overall, as $N-Z$ increases, the type of pairing does not suddenly switch from one to the other; rather, 
it gradually changes from spin-triplet to spin-singlet, a conclusion that survives from the $\Gamma$-less case. 
Note, however, that the correlation energies in all three cases have been reduced (though not sufficiently
to drown out the effects under study).

\subsection{Pairing gaps}
\label{sec:gaps}

Up to this point we have examined correlation energies, which reflect properties at the many-particle
level. It is worthwhile to investigate possible single-particle properties. One way to do so is by calculating 
the pairing gap from the odd-even mass differences. This can be arrived at via a simple difference formula: 
\begin{equation}
\Delta_0^{(3)}(n)=E(n)-\frac{1}{2} \left [ E(n-1)+E(n+1) \right] ,
\end{equation}
where $n$ is an odd neutron or proton number, with the other nucleon number set to an even value, and $E(n)$ is the ground state energy of the nucleus.

\begin{figure}[b]
\includegraphics[scale=0.5]{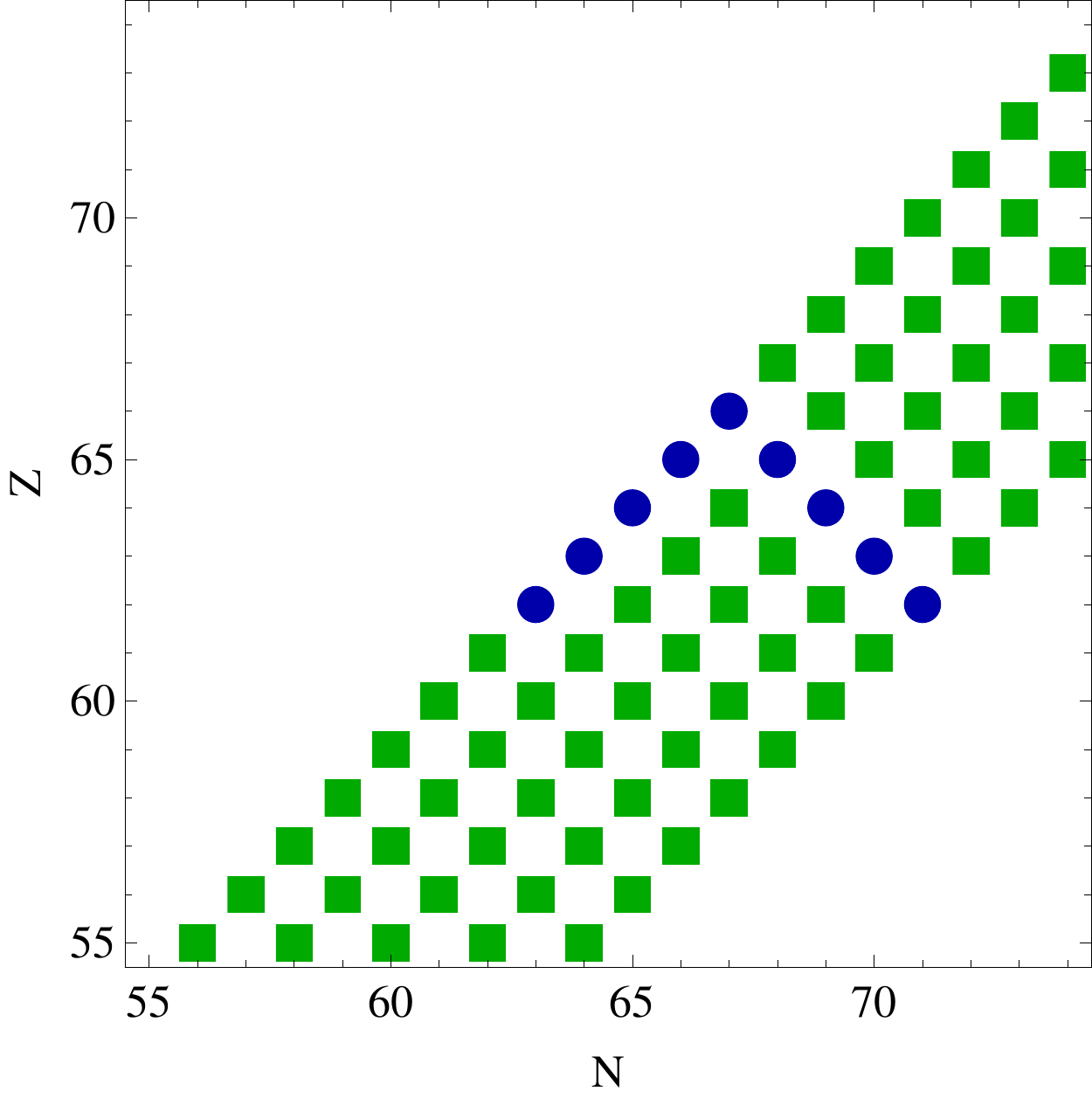}
\caption{(Color online) Pairing gaps for $55\leq N\leq 74$ and $N\geq Z$. Most nuclei have intermediate-size pairing gaps (green 
squares), though
two lines exhibit reduced gaps (blue circles).}
\label{fig_gaps}
\end{figure}

\begin{figure*}
\begin{minipage}[t]{0.32\linewidth}
\includegraphics[scale=0.29]{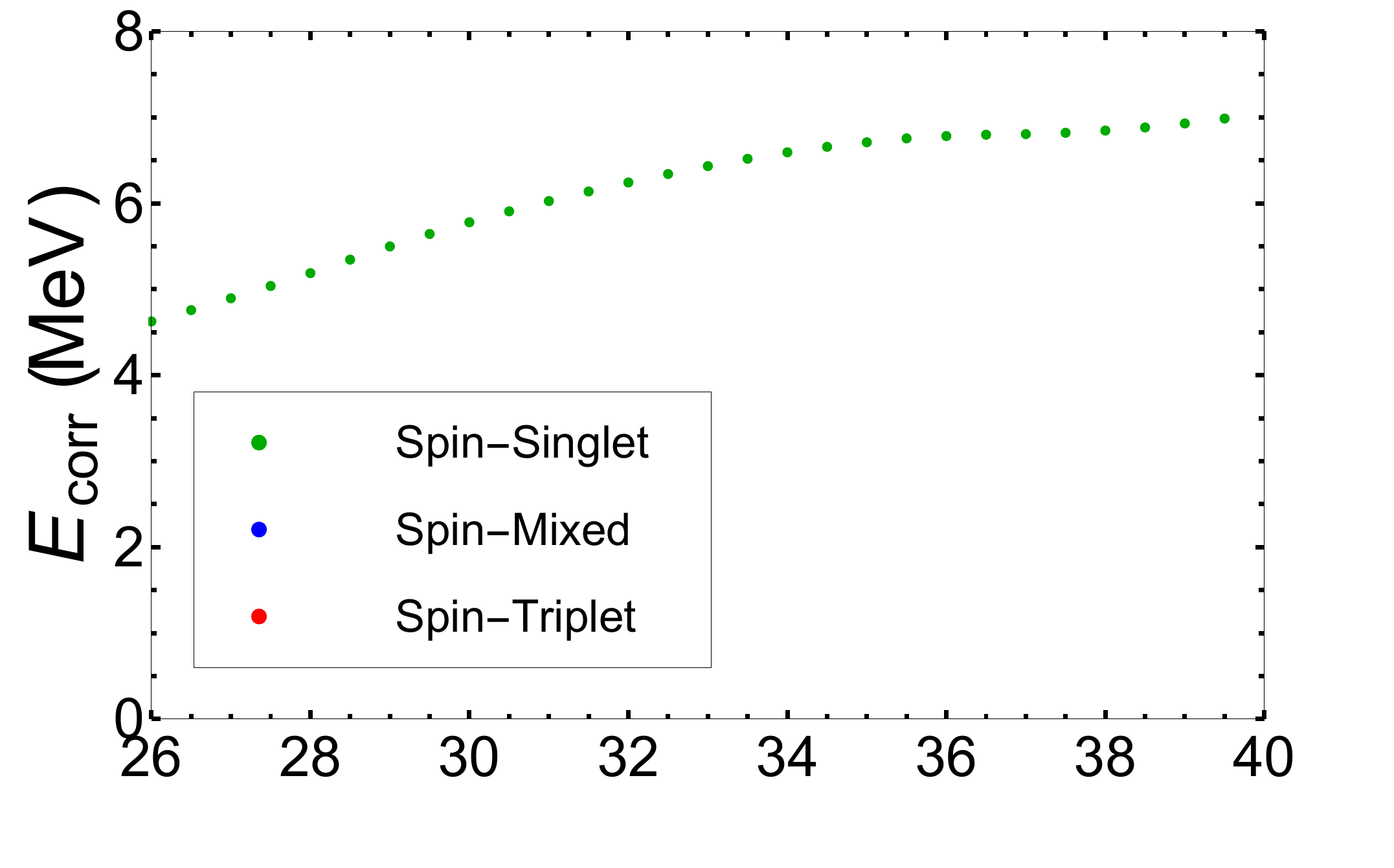}
\end{minipage}\hfill%
\begin{minipage}[t]{0.32\linewidth}
\includegraphics[scale=0.32]{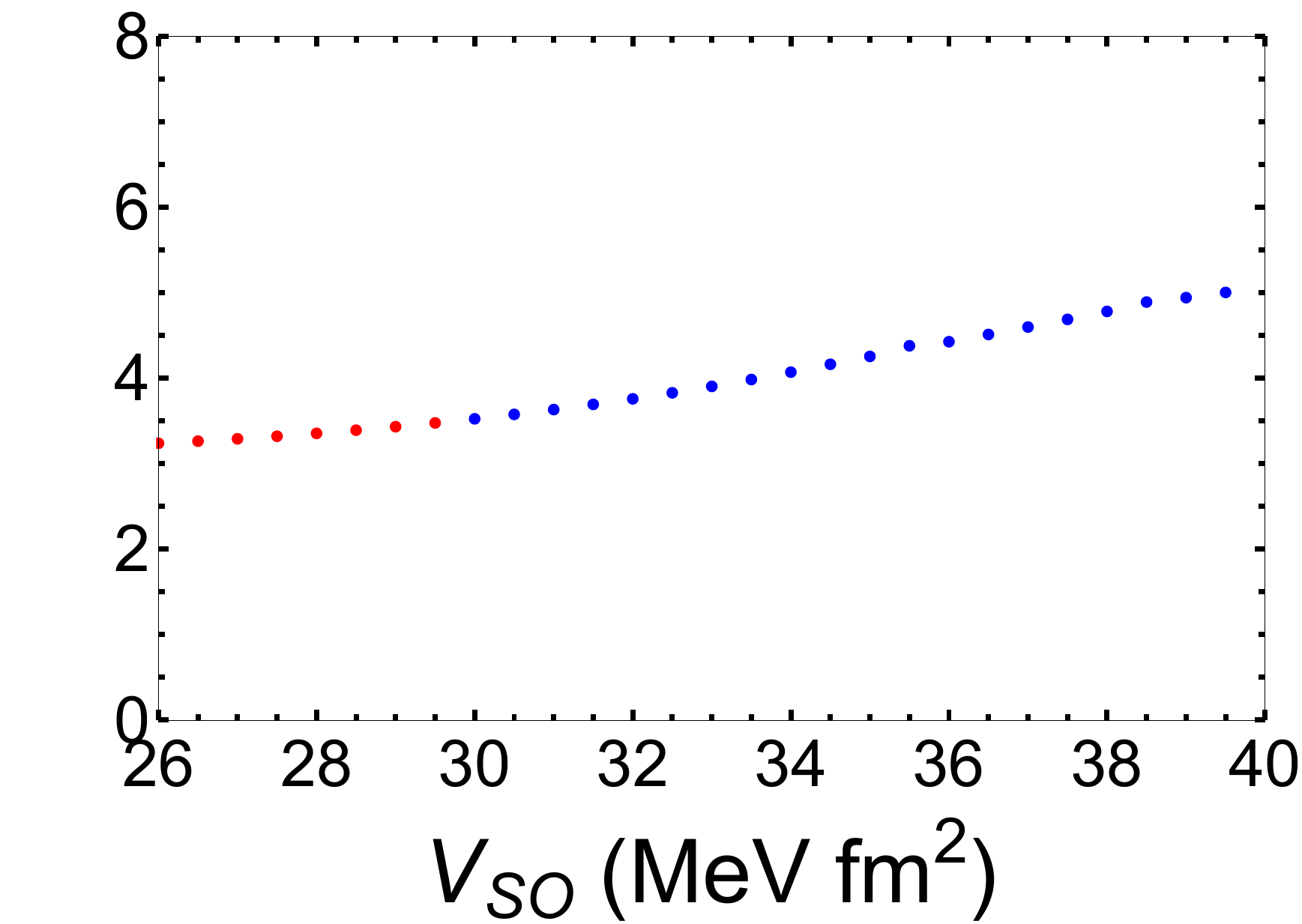}
\end{minipage}\hfill%
\begin{minipage}[t]{0.32\linewidth}
\includegraphics[scale=0.32]{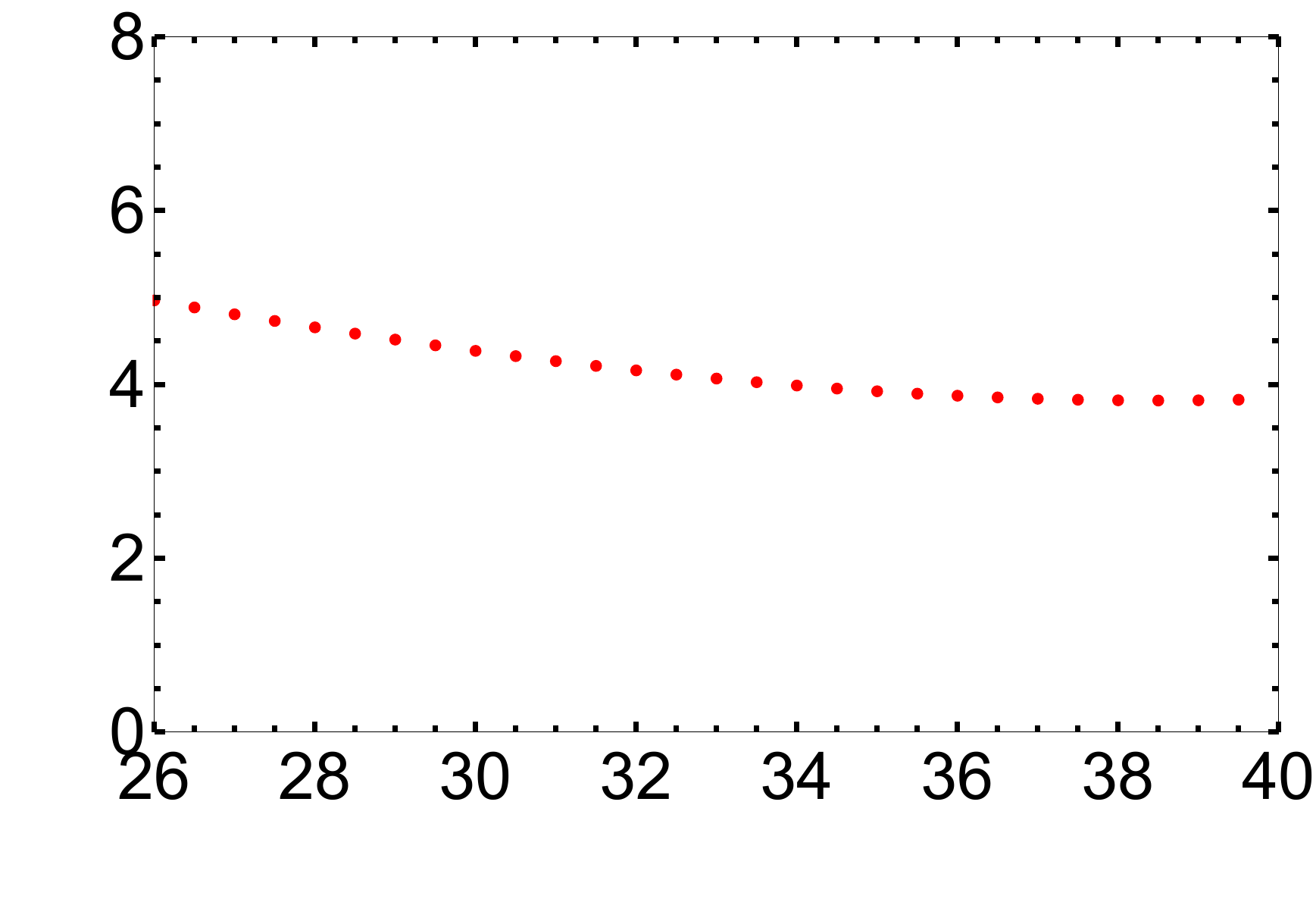}
\end{minipage}%
\caption{(Color online) Correlation energies (in MeV) vs. varying $V_{SO}$ (in MeV fm$^2$). These are plotted for the same nuclei as in Fig.~\ref{fig_trans1}: $^{132}_{60}$Nd in the left panel, $^{132}_{64}$Gd in the middle panel, and $^{132}_{66}$Dy in the right
panel. 
As in Fig.~\ref{pair_fig}, red indicates spin-triplet, green spin-singlet, and blue a mixed-state. \label{fig_VsoDep}}
\end{figure*}

The results are shown in Fig.~\ref{fig_gaps}. As in Ref.~\cite{GezerlisBertschLuo}, we find a signature of 
spin-triplet pairing in a chain of nuclei one unit away from $N=Z$ (above the proton drip line). What's different here is that
we also find a sequence of small pairing gaps on the $A=133$ line. This results from a more thorough search, imposing 
the odd-number parity per block across blocks, as well as the effects of the $\Gamma$ matrix. The placement of the $A=133$
line is not coincidental: it reflects the fact that above that line new states become available.
Additionally, since orbitals are included in the HFB space if their energies lie between the bounds $E_{\text{min}}$ and $E_{\text{max}}$, changing the energy bounds also changes the energy levels available for calculations for each mass number. We have checked the effects of increasing the $\Delta E$ and found no qualitative changes.

\subsection{One-body potential parameters}
\label{sec:one}
Our discussion of pairing gaps led to a study of the dependence (or lack thereof) 
of observed effects on specific model parameters. Continuing in that vein, we also varied the one-body and two-body
parameters appearing in Eqs.~(\ref{eqn_WSfunc}) and (\ref{eqn_interaction}), in order to determine the sensitivity
of the different pairing condensates to details of our input Hamiltonian model.

We start by \textit{ad hoc} varying the $V_{SO}$ one-body parameter, which controls the strength of the spin-orbit term.
Qualitatively, based on the ideas discussed in the Introduction, we expect that reducing $V_{SO}$ would lead to a strengthening
of spin-triplet pairing or, reversely, increasing $V_{SO}$ would lead to a suppression of spin-triplet pairing. Figure \ref{fig_VsoDep} shows the results of carrying out such calculations. Since our value for the spin-orbit strength above
was $V_{SO} = 33$ MeV fm$^2$, we here depart by 7 MeV fm$^2$ on either side.
The nuclei investigated are the same as those in Fig.~\ref{fig_trans1}; from left to right, they are $^{132}_{60}$Nd,  $^{132}_{64}$Gd, $^{132}_{66}$Dy. 

The results for $^{132}_{60}$Nd on the left show a fairly steady increase in correlation energy as $V_{SO}$ increases. The nature of the pairing also remains steadily spin-singlet. Additionally, we determined that 
the inclusion of the $\Gamma$ matrix once again causes a slight drop in the correlation energy, approximately 25\%.
The results for $^{132}_{66}$Dy on the right are similar, in that the spin-triplet character is not violated. Of particular interest is the decrease in correlation energy as $V_{SO}$ increases, rather than an increase. This reflects the qualitative
expectation that increased $V_{SO}$ is not favourable to spin-triplet pairing.
The results for $^{132}_{64}$Gd, shown in the middle of Fig.~\ref{fig_VsoDep}, are the most interesting. Similar to the $^{132}_{60}$Nd case, the correlation energy experiences an increase as $V_{SO}$ increases. But, also as $V_{SO}$ increases, the pairing nature changes from spin-triplet to a mixed-pairing state.

\begin{figure}[b]
\includegraphics[scale=0.45]{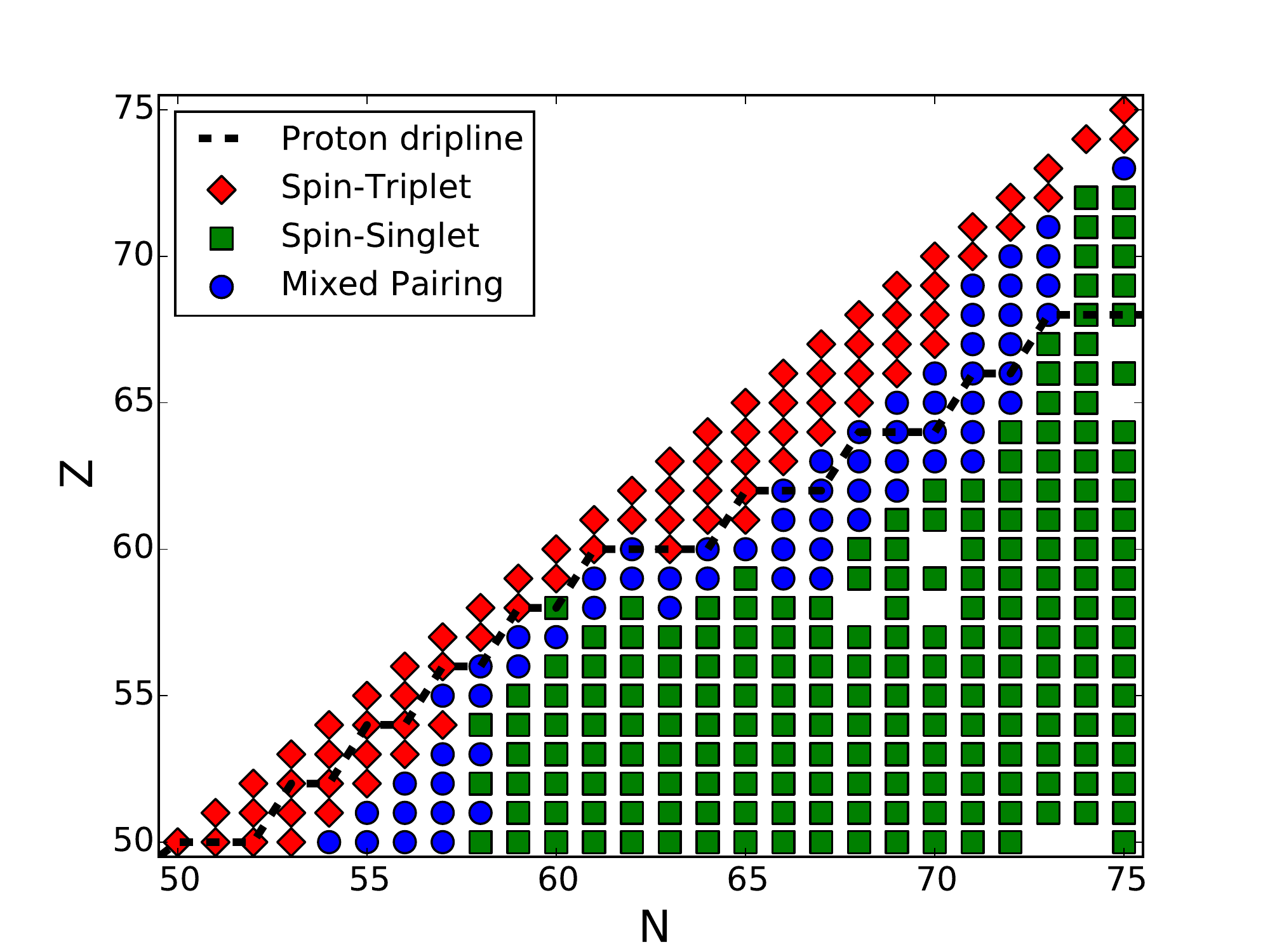}
\caption{(Color online) Chart of nuclides with $N \geq Z$ from $N=50$ to $N=75$ for $V_{SO}=0$. Notation is as in Fig.~\ref{pair_fig}. 
A vanishing spin-orbit strength leads to the appearance of many new spin-triplet and mixed-spin nuclei.
\label{pair_fig_Vso0}}
\end{figure}

From general
considerations one expects that the presence of the spin-orbit field (pushing
a partner to higher energies) suppresses pairing. This suppression is active
in both the isovector and isoscalar channels, see e.g. Fig. 11 in
Ref.~\cite{FrauendorfMacchiavelli}. On the other hand, the left panel of
Fig.~\ref{fig_VsoDep} shows the correlation energy increase as the spin-orbit
strength is increased. We have traced this to the orbitals selected
in the A=132 calculations, along with the relevant mixing. It's worth noting
that the trend exhibited by the correlation energy in $^{132}_{60}$Nd is analogous to
the behavior of the single-particle contribution to the total energy: the
spin-orbit field has a more pronounced effect here than in the mixed-spin or
spin-triplet nuclei.

Overall, Fig.~\ref{fig_VsoDep} supports the proposed suppressing nature of the nuclear spin-orbit interaction on spin-triplet pairing. 
Importantly, it shows that the spin-triplet and mixed-spin states, while depending on the overall magnitude of 
$V_{SO}$, do not result from a ``magic'' choice that is finely tuned: both exotic states appear for a large spectrum
of possible spin-orbit strength values.

The logical extension of the trends shown in Fig.~\ref{fig_VsoDep} is to examine what would happen if in our calculations
we switched off the spin-orbit strength, $V_{SO}$, completely. This is, of course, unphysical, so functions only as a 
consistency check. The results are shown in Fig.~\ref{pair_fig_Vso0}: we find clear signatures of spin-triplet pairing
across the entire $N=Z$ line, as could be expected from the qualitative arguments given in the Introduction, as well
as slightly off it. This figure
also nicely illustrates the other cause behind the suppression of spin-triplet pairing: the isospin asymmetry gradually
turns spin-triplet states into mixed-spin states (further below the $N=Z$ line) which eventually turn into spin-singlet
states. 

In the spirit of more carefully probing the input Hamiltonian parameters, we have also considered the possibility
that these may change according to how neutron-rich the nucleus under study is. In Ref.~\cite{GezerlisBertschLuo} the 
parameters $V_{WS}$ and $V_{SO}$ were kept constant regardless of the value of $N-Z$ or $A$. Here, motivated by a
standard formula,\cite{BohrMottelson} we also examine the effect of changing these parameters 
(individually or in concert) on 
the correlation energies for $^{132}_{60}$Nd, $^{132}_{64}$Gd, and $^{132}_{66}$Dy. In keeping with our default values,
we explore the following dependences:
\begin{align}
V_{WS}=&\left(-50+33\frac{N-Z}{A}\right) ~\text{MeV}, \nonumber \\
V_{SO}=&r_0^2\left(22-15\frac{N-Z}{A}\right) ~\text{MeV},
\label{eq:Bohr}
\end{align}
where $r_0=1.27$ fm. The results are shown in Table \ref{table_VwsVso}.

\begin{table}[t]
\begin{tabular}{|c|c|c|c|c|}
\hline
Nucleus & $V_{WS}$ & $V_{SO}$ & $E_{\text{corr}}$ & singlet fraction\\
\hline
$^{132}_{60}$Nd & -50 & 33 & 6.432 & 1\\
 & -50 & 33.28 & 6.482 & 1\\
 &-47 & 33.28 & 8.016 &1\\
\hline
$^{132}_{64}$Gd &-50&33&3.904&0.2784\\
&-50&34.75&4.209&0.3027\\
&-49&34.75&4.138&0.3096\\
\hline
 $^{132}_{66}$Dy&-50&33&4.067&0\\
  &-50&35.48& 3.894 &0\\
 \hline
\end{tabular}
\caption{Correlation energies for $^{132}_{60}$Nd, $^{132}_{64}$Gd, and $^{132}_{66}$Dy with various $V_{WS}$ and $V_{SO}$. Also shown is the ``singlet fraction'', namely
the ratio of the spin-singlet amplitude and the total pairing amplitude.}
\label{table_VwsVso}
\end{table}

First, we note that the singlet fraction (defined as the ratio of spin-singlet and total pairing amplitudes) is not 
impacted by these alternative choices for $V_{WS}$ and $V_{SO}$: $^{132}_{60}$Nd remains firmly on the spin-singlet side, 
$^{132}_{64}$Gd still exhibits mixed-spin pairing, and $^{132}_{66}$Dy is still overwhelmingly a spin-triplet nucleus
(of course, the main point of Eq.~(\ref{eq:Bohr}) was to explore an $N-Z$ dependence, which would not show up for 
$^{132}_{66}$Dy). Next, we notice that changing the $V_{SO}$ parameter alone by these small amounts does not have an appreciable effect on the correlation energy or on the nature of the pairing condensate: this is hardly surprising, 
since precisely the same modification was carried out for the purposes of Fig.~\ref{fig_VsoDep}. Moving on to 
the lines of the table showing a change of both $V_{WS}$ and $V_{SO}$: $^{132}_{60}$Nd shows an increased
correlation energy, though no fundamental change in the nature of the pairing, while $^{132}_{64}$Gd is barely impacted.

\subsection{Pairing strength variation}
\label{sec:two}

\begin{figure*}[t]
\begin{tabular}{c}
\includegraphics[scale=0.2]{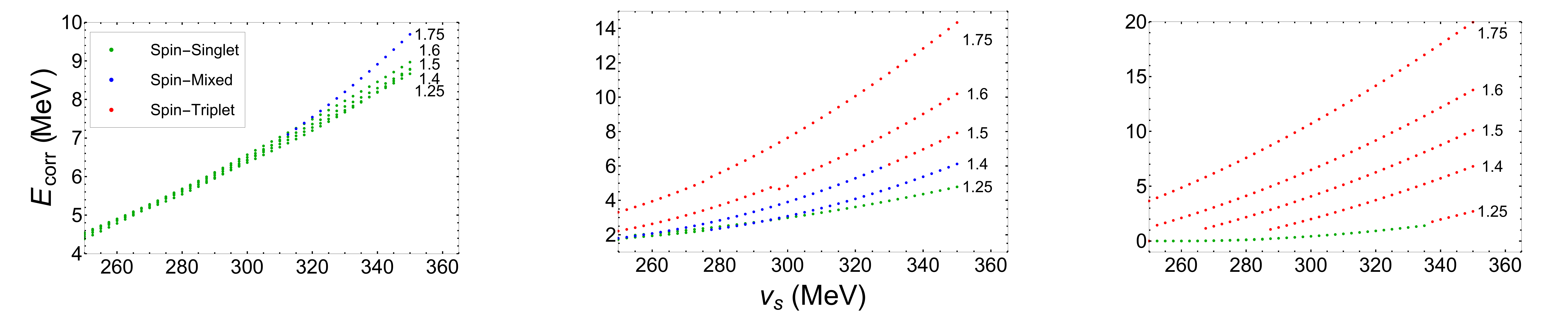}
\end{tabular}
\caption{(Color online) Correlation energies (in MeV) vs. varying $v_{s}$, with $v_t/v_s$ ratios of 1.25, 1.4, 1.5, 1.6, and 1.75. On the left is $^{132}_{60}$Nd, on the right is $^{132}_{66}$Dy, and in the middle is $^{132}_{64}$Gd. All plots indicate a change from spin-singlet (green) towards spin-triplet (red) possibly with an intermediate mixed-state (blue) as the $v_t/v_s$ ratio increases, and an increase in correlation energy as $v_s$ increases. \label{fig_VsVt}}
\end{figure*}

As noted in section~\ref{sec:Ham}, in Ref.~\cite{BertschLuo} the pairing strengths $v_s$ and $v_t$ 
(for spin-singlet and spin-triplet, respectively)
were arrived at by comparing with phenomenological shell-model Hamiltonians. The $v_s$ was estimated to be $\sim 280$ and
the ratio between the two parameters 
was found to be $\sim 1.65$, so the two parameter values were conservatively estimated to be 
300 MeV and 450 MeV, respectively. These were also the strengths used in Ref.~\cite{GezerlisBertschLuo} and 
in the present work up to this point. Here, once again, 
we attempt to see if there is a sensitive dependence on
the specific values chosen or, alternatively, if 
the conclusions on the pairing condensate character(s) are fairly robust.
We have examined $v_s$ values from 250 to 350 MeV, for $v_t/v_s$ ratios of 1.25, 1.4, 1.5, 1.6, and 1.75.
The results for the correlation energies are shown in Fig.~\ref{fig_VsVt}. 
As in earlier figures, 
on the left is $^{132}_{60}$Nd, on the right is $^{132}_{66}$Dy, and in the middle is $^{132}_{64}$Gd. 

An unmistakeable
trend is that as we move to the right of the $x$-axis in each figure, 
increasing both $v_s$ and $v_t$, the correlation energy also increases,
a natural result of having stronger interactions. Note that for $^{132}_{60}$Nd the spread of the results for 
different $v_t/v_s$ ratios is relatively small: for most strengths and ratios the correlation energy is approximately
the same. Even for this clearly spin-singlet paired nucleus, a very large $v_t/v_s$ ratio starts to lead to mixed-spin
pairing. The behavior of $^{132}_{64}$Gd and $^{132}_{66}$Dy is quite different: for these nuclei different $v_t/v_s$ ratios
lead to considerably different correlation energies. For small ratios both nuclei exhibit spin-singlet pairing, while overall
$^{132}_{64}$Gd spans the gamut from spin-singlet, to mixed-spin, to spin-triplet pairing: the presence of mixed-spin
pairing for reasonable values of the pairing strengths is fairly well established here. In the case of 
$^{132}_{66}$Dy, we see that for small pairing strength ratios (at most values of $v_s$) 
there is no pairing whatsoever (zero correlation energy), while for any other ratio we find a clean signal corresponding
to spin-triplet.

\section{Summary and Conclusion}

In summary, we have improved the solution of the Hartree-Fock-Bogoliubov problem in comparison to earlier work, by
employing the gradient method and explicitly including the Hartree-Fock $\Gamma$ field. We have then carried out 
calculations for nuclides with $N \geq Z$ for $N=50$ to $N=75$. These were carried out by constraining the average
proton and neutron particle numbers. We find that including $\Gamma$ leads to a suppression of the correlation energy,
which is not, however, sufficient to eliminate the mixed-spin pairing condensate for a number of nuclei below the
proton-drip line. We have also taken the opportunity to examine the particle number fluctuations, which end up being
reasonably small and closely following the magnitude of the correlation energy. Furthermore, we have investigated the 
effect of artificially constraining the spin-singlet and spin-triplet pairing amplitudes away from the ground-state minima:
the results are qualitatively unchanged. Regarding the pairing gaps, we found a sequence of spin-triplet character (above the
drip line) as well as a line of reduced pairing gaps below the proton drip line. We have also attempted to modify our
model Hamiltonian one-body and two-body parameters. We found that the spin-orbit strength quenches the 
spin-triplet pairing, though not dramatically so for values close to our default ones. Similarly, the pairing strengths
impact the nature of the pairing condensate, but we see no signs of our default values being finely tuned. 

In terms of future work, the natural next step would be to extend our mean-field theory so that it can handle broken
symmetries, like particle number and angular momentum. \cite{BertschRobledo} For example, at this stage we have not 
made statements on the ground-state spins for the nuclei that we find exhibit mixed-spin pairing. Such an extension would
allow one to make predictions for spectroscopic quantities as well as two-particle transfer reactions. Another natural 
avenue of future work would be to address nuclear deformation \cite{Gambacutra}: while the prediction of mixed-spin pairing for experimentally
accessible nuclei is tantalizing, the results we have produced have been until now restricted to the case of spherical 
symmetry, which is known not to apply to the region of interest. The pairing would certainly be weakened in a fully 
three-dimensional Hartree-Fock-Bogoliubov calculation, though it remains to be seen by how much and in what way.

\

\section*{Acknowledgments}
The authors acknowledge insightful discussions with G.~F. Bertsch, 
A.~O. Macchiavelli, and C.~E. Svensson. 
This work was supported 
by the Natural Sciences and Engineering Research Council (NSERC) of Canada
and the Canada Foundation for Innovation (CFI). 
Computations were performed at NERSC and at SHARCNET.

\end{document}